# Monte Carlo Simulations of Macho Parallaxes From a Satellite

**Thomas Boutreux**[1]

CEA, DSM, DAPNIA, SPP

Centre d'Etudes de Saclay, 91191 Gif-sur-Yvette, France

and

**Andrew Gould**[2]

Department of Astronomy, Ohio State University

Columbus, OH 43210

boutreux@clipper.ens.fr, gould@payne.mps.ohio-state.edu

---

**Abstract**

Three ongoing microlensing experiments have found more candidate events than expected from the known stars. These experiments measure only one parameter of the massive compact halo objects (machos), the magnification time scale of the events. More information is required to understand the nature of the machos. A satellite experiment has been proposed to measure their projected transverse speed $\tilde{v} = v/(1-z)$, where $v$ is the macho transverse speed and $z$ its distance divided by the distance of the source. Measurement of $\tilde{v}$ would determine whether the machos were in the Galactic disk, Galactic halo, or in the Large Magellanic Cloud (LMC). We simulate events observed toward the LMC by the Earth and by a satellite in an Earth like heliocentric orbit. To leading order, such an experiment determines $\tilde{v}$ up to a two fold degeneracy. More precise measurements break the degeneracy. We show that with photometric precisions of 3% to 4% and approximately 1 observation per day, $\tilde{v}$ can be measured with a maximum error of 20% for 70% to 90% of events similar to the ones reported by the EROS and MACHO collaborations. The projected transverse velocity is known with the same maximum error for 60% to 75% of these events. This 20% maximum error is not a 1 $\sigma$ error but is mostly due to degeneracy between two possible solutions, each one being localized to much better than 20%. These results are obtained with an Earth-satellite separation of 1 AU, and are improved by a larger separation.

# 1. Introduction

The EROS collaboration (Aubourg et al. 1993; Aubourg et al. 1995a, 1995b) and the MACHO collaboration (Alcock et al. 1993, 1995b) have reported to date a total of five candidate microlensing events of stellar sources in the Large Magellanic Cloud (LMC), more than predicted from the known luminous stars of the Milky Way (Bahcall et al. 1994; Gould, Bahcall & Flynn 1995) and the LMC (Gould 1995b), but less than expected with a standard spherical halo (Paczyński 1986; Griest 1991). The MACHO collaboration (Alcock et al. 1995a; Bennett et al. 1995) and OGLE collaboration (Udalski et al. 1994) have reported several dozen candidate events toward the Galactic bulge, also more than expected from known stars (Paczyński et al. 1994; Han & Gould 1995; Zhao, Spergel, & Rich 1995).

To understand both the nature of the massive compact halo objects (machos) which generate these events and the structure of the Galaxy, one must collect as much information as possible about each event. The current microlensing searches yield only one characteristic of the lens, the time scale $t_e$ of the event: $t_e = r_e/v$, where $v$ is the magnitude of the transverse velocity $\mathbf{v}$ of the macho relative to the Earth–source line of sight, and $r_e$ is the Einstein radius of the macho:

$$r_e^2 = \frac{4G}{c^2} d_{source} M z (1-z). \qquad (1.1)$$

Here M is the macho mass, $d_{source}$ is the distance between the Earth and the source (in e.g. the LMC) and $z$ is the ratio of the macho distance $d_{lens}$ to the source distance:

$$z = \frac{d_{lens}}{d_{source}}. \qquad (1.2)$$

A variety of methods of obtaining additional information have been discussed. The most efficient and promising one seems to be the observation of the same events by both Earth-based and space-based telescopes (Refsdal 1966; Gould 1994, 1995a). This method uses the fact that the event looks substantially different as seen from the two locations to measure two new characteristics of the macho, the two components of the "projected velocity" $\tilde{\mathbf{v}}$, the transverse velocity projected onto the observer plane,

$$\tilde{\mathbf{v}} = \mathbf{v}\frac{d_{source}}{d_{source} - d_{lens}} = \frac{\mathbf{v}}{1-z}. \tag{1.3}$$

Table 1 gives the value of the projected speed for the different possible populations of machos on lines of sight toward the LMC (disk, thick disk, Galactic halo, LMC halo, LMC disk). It shows that by measuring the projected speed of each event with an error of $\lesssim 20\%$, one can determine the population to which each macho belongs with good confidence. Thus, for example, one could determine whether the machos make up a significant fraction of the dark halo. Moreover, the additional measured parameters constrain the kinematics and the mass spectrum of whatever population is detected.

TABLE 1

Mean Projected Velocities

| Component | $\langle v \rangle$ km s$^{-1}$ | $z$ | $\langle \tilde{v} \rangle$ km s$^{-1}$ |
|---|---|---|---|
| Galactic Disk | 50 | 0.01 | 50 |
| Galactic Thick Disk | 100 | 0.04 | 105 |
| Galactic Halo | 220 | 0.20 | 275 |
| LMC Halo | 100 | 0.94 | 1700 |
| LMC Disk | 30 | 0.99 | 3000 |

By comparing the peak magnifications and peak times of the event as seen from the Earth and satellite, one could determine $\tilde{\mathbf{v}}$ up to a four fold degeneracy and $\tilde{v}$ up to a two fold degeneracy (Refsdal 1966; Gould 1994). Gould (1995a) showed that these degeneracies can in principle be broken by measuring the small difference in the event time scales due to the relative motion of the Earth and the satellite. He also gave a rough estimate of the precision required to break the degeneracy. Here we report on Monte Carlo simulations of observations of the same microlensing events from the ground and a satellite. We have determined the conditions needed to break the degeneracy for a large fraction of the events and which would permit measurement of the projected speed $\tilde{v}$ with an error $\lesssim 20\%$. We focus attention on the photometric precisions, the frequency of the observations, and the Earth–satellite separation. The simulations consider a broad range of possible macho masses. We restrict our study to observations toward the LMC. Observations toward the Galactic bulge are physically and geometrically different, and will be discussed elsewhere (Gaudi & Gould 1995).

## 2. Error Analysis

The mathematical analysis underlying the Monte Carlo is substantially more involved than is typically the case. We therefore include in this section a complete mathematical description. The reader who is interested primarily in the results can skip directly to § 3. However, as we discuss at the end of this section, most of these results can be understood analytically and therefore study of the present section is useful for a physical as well as a mathematical understanding of the problem.

The magnification $A$ of a microlensed star is a function only of the separation $x$ of the lens and star in units of the angular Einstein radius $\theta_e = r_e/d_{lens}$. Explicitly $A(x) = (x^2 + 2)/x(x^2 + 4)^{1/2}$ (Paczyński 1986). Assuming rectilinear motion of the observer, source, and lens, $x$ evolves with time as

$$x(t) = \sqrt{\omega^2(t-t_0)^2 + \beta^2}, \tag{2.1}$$

where $\omega = t_e^{-1}$, $\beta$ is the impact parameter in units of the Einstein radius, and $t_0$ is the time of maximum magnification. The flux $F(t)$ from the microlensed star is therefore a function of five parameters $a_i = (t_0, \beta, \omega, F_0, B)$,

$$F(t; t_0, \beta, \omega, F_0, B) = F_0 A[x(t; t_0, \beta, \omega)] + B, \tag{2.2}$$

where $F_0$ is the unlensed flux and $B$ is the light from any unresolved additional sources that are not lensed (such as the lens itself or a binary companion to the lensed star). If a set of measurements $y_k$ are made at times $t_k$, then the parameters may be fit by minimizing $\chi^2 = \sum_k [y_k - F(t_k; a_i)]^2 / \sigma_k^2$ where $\sigma_k$ is the error in the $k$th measurement. By differentiating $\chi^2$ in the neighborhood of the solution one finds $c_{ij}$, the covariance matrix of the $a_i$,

$$c \equiv b^{-1}; \qquad b_{ij} = \sum_k \sigma_k^{-2} \frac{\partial F(t_k; a_1...a_5)}{\partial a_i} \frac{\partial F(t_k; a_1...a_5)}{\partial a_j}. \tag{2.3}$$

If measurements of the event are also made from a satellite, then the flux can be described by five additional parameters $a_i' = (t_0', \beta', \omega', F_0', B')$. (In the simulations described in the following section, we assume that the errors are given by $\sigma_k = \sigma_E F_0 [A(t_k)]^{1/2}$ and $\sigma_k' = \sigma_S F_0' [A'(t_k)]^{1/2}$ where $\sigma_E$ and $\sigma_S$ are fixed parameters.) The covariance matrix for both sets of measurements combined is then a $10 \times 10$ block diagonal matrix with the upper block given by equation (2.3) and the lower block given by a similar equation for the satellite. There are now 10 parameters $a_i$.

If the Earth and the satellite telescopes have the same filters, then $F_0 = F_0'$. Formally, this constraint can be written

$$\alpha_i a_i = 0, \tag{2.4}$$

where $\alpha_i = (0, 0, 0, 1, 0, 0, 0, 0, -1, 0)$ and where we have introduced the Einstein summation convention. To see how to impose the constraint, assume first that it has a finite but very small error $Q$: $\alpha_i a_i = 0 \pm Q$. Then the associated inverse covariance matrix is $b_{ij}^Q = \alpha_i \alpha_j / Q^2$. The restricted covariance matrix is then $\tilde{c} = (b + b^Q)^{-1}$ (see e.g. Gould 1989). Letting $Q \to 0$ yields,

$$\tilde{c}_{ij} = c_{ij} - \frac{c_{il}\alpha_l c_{jk}\alpha_k}{c_{mn}\alpha_m \alpha_n}. \tag{2.5}$$

If the filters are the same, the unlensed light is also equal, $B = B'$, and this leads to a similar constraint with $\alpha_i = (0, 0, 0, 0, 1, 0, 0, 0, 0, -1)$. Equation (2.5) should then be applied twice successively, once for each constraint.

To a first approximation, the Earth and the satellite are at rest with respect to each other, so that $\omega \simeq \omega'$. As discussed by Gould (1994), the projected velocity is then given by

$$\tilde{\mathbf{v}} = \omega L \frac{\Delta \mathbf{x}}{(\Delta \mathbf{x})^2}, \qquad \Delta \mathbf{x} \equiv (\omega \Delta t, \Delta \beta). \tag{2.6}$$

Here $L$ is the magnitude of $\mathbf{L}$, the projection of the Earth-satellite separation vector onto the plane of the sky, $\Delta t \equiv t'_0 - t_0$, and $\Delta \beta$ is the difference in impact parameters. The components of $\Delta \mathbf{x}$ are respectively parallel and perpendicular to the projected separation vector. There is a two fold degeneracy in the magnitude $|\Delta \beta| = \Delta \beta_\pm$ depending on whether the source is on the opposite or same side of the macho as seen from the Earth and satellite. There is a further two fold degeneracy in the sign of $\Delta \beta$ depending on the relative orientation of the source-lens separation as seen from the two observers. However, these degeneracies can be broken by measuring the small difference $\Delta \omega = \omega' - \omega$ due to the relative Earth-satellite motion. This results in a constraint

$$u_\parallel \omega \Delta t + u_\perp \Delta \beta - L \Delta \omega = 0, \tag{2.7}$$

where $u_\parallel$ and $u_\perp$ are the components of the velocity of the satellite relative to the Earth parallel and perpendicular to the projected satellite-Earth separation vector (Gould 1995a).

When constructing the simulations, we of course know which side of the macho the source passes as seen from the Earth and the satellite and hence we know not only the magnitudes of $\beta$ and $\beta'$, but also the magnitude and sign of $\Delta\beta$. However, we initially assume that we measure only the magnitudes of $\beta$ and $\beta'$ and that all four possible values of $\Delta\beta$ are equally likely. We then test these possible solutions separately. Consider for example the $+\Delta\beta_-$ solution, which we label "$-+$". We first form the 3-vector of parameters $a_i^{-+} = (\Delta t, +\Delta\beta_-, \Delta\omega)$. We then calculate the $3\times 3$ covariance matrix of these quantities, $c_{ij}^{-+} = \tilde{c}_{i,j} - \tilde{c}_{i,j+5} - \tilde{c}_{i+5,j} + \tilde{c}_{i+5,j+5}$. Next we formulate the constraint as in equation (2.4) with $\alpha_i = (u_\parallel \omega/L, u_\perp/L, -1)$ which leads to an adjustment of the covariance matrix $c^{-+} \to \tilde{c}^{-+}$ [see eq. (2.5)] and an adjustment of $a_i^{-+}$ (see e.g. Gould 1989),

$$\tilde{a}_i^{-+} = \tilde{c}_{ij}^{-+}(b_{jk}^{-+}a_k^{-+}) = a_i^{-+} - \frac{\alpha_l a_l^{-+}}{c_{mn}^{-+}\alpha_m\alpha_n} c_{ij}^{-+}\alpha_j, \qquad (2.8)$$

where $b^{-+} \equiv (c^{-+})^{-1}$. We evaluate the $\chi^2$ of this best $(-+)$ solution given the constraint:

$$\chi^2_{-+} = (\tilde{a}_i^{-+} - a_i^{-+})b_{ij}^{-+}(\tilde{a}_j^{-+} - a_j^{-+}) = \frac{(\alpha_l a_l^{-+})^2}{c_{mn}^{-+}\alpha_m\alpha_n}. \qquad (2.9)$$

At this point in the analysis, we assume that the best estimates of $\Delta t$, $\Delta\omega$, $\beta$ and $\beta'$ are equal to the true values. (We consider the effect of possible differences between the best estimates and the true values below, but as we eventually show, these are relatively unimportant.) For the true values and for the right choice of $\Delta\beta$, we have $\alpha_l a_l = 0$. Since the four possible choices of $a_l$ differ only in the second ($\Delta\beta$) component, $\chi^2$ can be written

$$\chi^2_{-+} = \frac{(\alpha_2 \delta\beta^{-+})^2}{c_{mn}^{-+}\alpha_m\alpha_n}, \qquad (2.10)$$

where $\delta\beta^{-+} = \Delta\beta - (+\Delta\beta_-)$ is the difference between the true and trial values of $\Delta\beta$. For example, if the true solution is $\Delta\beta = -\Delta\beta_-$ then $\delta\beta^{++} = 2\beta'$, $\delta\beta^{-+} = 2(\beta' - \beta)$, $\delta\beta^{--} = 0$, and $\delta\beta^{+-} = -2\beta$.

This analysis is repeated for all four solutions and $\chi^2$ is evaluated for each. The only important modification is that for the $\pm\Delta\beta_+$ solutions, the $\Delta\beta$ components of the covariance matrix are $c_{i2}^{+\pm} = c_{2i}^{+\pm} = \pm(-\tilde{c}_{2,i} + \tilde{c}_{2,i+5} - \tilde{c}_{7,i} + \tilde{c}_{7,i+5})$ ($i \neq 2$) and $c_{22}^{+\pm} = \tilde{c}_{2,2} + \tilde{c}_{2,7} + \tilde{c}_{7,2} + \tilde{c}_{7,7}$, while for the $-\Delta\beta_-$ solutions, the ($i \neq 2$) components are $c_{i2}^{--} = c_{2i}^{--} = -c_{i2}^{-+}$.

In the simulations, we accept solutions if $\chi^2 \leq 9$. We also calculate the intrinsic error in each of the allowed solutions. The fractional error in the vector solution is defined as $\langle(\delta\tilde{\mathbf{v}})^2\rangle^{1/2}/\tilde{v}$ which for small values can be approximated as $\langle(\delta\Delta\mathbf{x})^2\rangle^{1/2}/\Delta x$, where we recall that $\Delta\mathbf{x} \equiv (\omega\Delta t, \Delta\beta) = (a_1, a_2)$. To calculate this quantity, we apply the constraint (2.7) to e.g. $c^{-+}$ and form $\tilde{c}^{-+}$ [see eq. (2.5)] and then compute

$$\frac{\langle(\delta\tilde{\mathbf{v}})^2\rangle^{1/2}}{\tilde{v}} = \left[\frac{\tilde{c}_{11}^{-+} + 2\tilde{c}_{12}^{-+} + \tilde{c}_{22}^{-+}}{(a_1^{-+})^2 + (a_2^{-+})^2}\right]^{1/2}. \tag{2.11}$$

The fractional error in the scalar solution is defined as $\langle(\delta\tilde{v})^2\rangle^{1/2}/\tilde{v}$ which can be approximated as $\langle(\delta\Delta x)^2\rangle^{1/2}/\Delta x \simeq (1/2)\langle[\delta(\Delta x)^2]^2\rangle^{1/2}/(\Delta x)^2$. After some algebra, we find

$$\frac{\langle(\delta\tilde{v})^2\rangle^{1/2}}{\tilde{v}} = \frac{\left(\sum_{i,j=1}^{2} \tilde{c}_{ij}^{-+} a_i^{-+} a_j^{-+}\right)^{1/2}}{(a_1^{-+})^2 + (a_2^{-+})^2}. \tag{2.12}$$

We can give some further insight into the formalism derived in this section by restricting consideration to the case at hand: observations toward the LMC. Since the LMC is very close to the ecliptic pole, the constraint can be very well approximated by $\alpha = (0, \Omega, 1)$ where $\Omega = 2\pi$ yr$^{-1}$ is the frequency of the Earth's orbit. Then equation (2.10) can be written as

$$\chi^2 = \left(\frac{\Omega t_e \delta\beta}{\sigma_\omega}\right)^2 \left[1 + \left(\frac{\Omega t_e \sigma_\beta}{\sigma_\omega}\right)^2\right]^{-1}, \tag{2.13}$$

where $\sigma_\omega = c_{33}^{1/2} t_e$ is the dimensionless error in the time-scale difference, $\sigma_\beta = c_{22}^{1/2}$, and where we have for simplicity assumed that the correlation coefficient is small.

For typical parameters ($\beta, \beta' \sim 0.5$), one finds $\sigma_{\beta,-}/\sigma_\omega \sim 0.5$ for the $\pm\Delta\beta_-$ solutions and $\sigma_{\beta,+}/\sigma_\omega \sim 8$ for the $\pm\Delta\beta_+$ solutions. As discussed by Gould (1995a), the higher errors in the latter case arise from the strong correlation in the errors of $\beta$ and $\beta'$. This correlation is induced by the constraints $F_0 = F_0'$ and $B = B'$.

Consider first the $\pm\Delta\beta_-$ cases. For $t_e \lesssim \Omega^{-1}\sigma_\omega/\sigma_\beta \sim 116\,\mathrm{days}$,

$$\chi^2_{-\pm} = \left(\frac{\Omega t_e \delta\beta^{-\pm}}{\sigma_\omega}\right)^2. \tag{2.14}$$

One might in principle also consider the opposite regime. First, however, such long events are not observed. Second, even if they were observed, they would be susceptible to ground-based parallax (Gould 1992; Alcock et al. 1995c) so that space-based parallax would not be necessary. Thus, equation (2.14) holds for all relevant cases. The situation is quite different for the $\pm\Delta\beta_+$ cases. Here

$$\chi^2_{+\pm} = \left(\frac{\delta\beta^{+\pm}}{\sigma_{\beta,+}}\right)^2, \qquad (t_e \gtrsim 7\,\mathrm{days}), \tag{2.15}$$

so that the degeneracy breaking is independent of time scale for all but the shortest events.

Using equations (2.14) and (2.15), one can gain a good idea of the sensitivity of the entire experiment. First, note that for fixed $\beta$ and $\beta'$, and for fixed number of observations per Einstein radius crossing time, the errors in $\Delta\mathbf{x} = (\omega\Delta t, \Delta\beta)$ and in $\Delta\omega/\omega$ are also fixed. The distributions of $\beta$ and $\beta'$ depend only on the ratio $d_{sat}/\tilde{r}_e$ of the Earth-satellite separation to the projected Einstein ring. Hence, if one ignores the problem of degeneracy breaking, the sensitivity of the experiment depends only on this ratio and not on the time scales of the events. The principal problem in degeneracy breaking is to separate the $\Delta\beta_+$ from the $\Delta\beta_-$ solutions. If this can be done, it is possible to determine $\tilde{v}$. If the true solution is $\Delta\beta_-$ then the $\Delta\beta_+$ solutions must be eliminated and visa versa. Clearly, the former is more difficult because the $\Delta\beta_+$ errors are larger. Thus, the overall requirements of the experiment are set by equation (2.15). Notice that this is also independent of $t_e$ (provided $t_e \gtrsim 7$ days). The smaller value of $\min|\delta\beta^{+\pm}| = 2\min\{\beta,\beta'\} \sim \mathcal{O}(1)$. It can of course be less than unity, but if it is too much less, then the $\Delta\beta_\pm$ solutions become so similar that it is not important to break the degeneracy.

Hence, one arrives at a basic requirement $\sigma_{\beta,+} \lesssim 0.1$ to routinely break the principal degeneracy at the $3\,\sigma$ level. This implies $\sigma_{\beta,-} \lesssim 0.01$. If this requirement is met, then it follows immediately that the intrinsic errors around each solution are unimportant compared to the degeneracy breaking errors provided $d_{sat}/\tilde{r}_e \gtrsim 0.05$. Thus one expects the experiment to attain a maximum sensitivity over a fairly broad range from $d_{sat} \lesssim \tilde{r}_e$ (beyond which the satellite falls outside the Einstein ring) and $d_{sat} \gtrsim \tilde{r}_e/20$ (beyond which the intrinsic errors dominate and rise as $\tilde{r}_e^{-1}$).

Finally, one can see that if the experiment is designed primarily to break the (two fold) speed degeneracy, then the (four fold) velocity degeneracy will usually also be broken. The most difficult case is $\pm\Delta\beta_-$. For definiteness, take $+\Delta\beta_-$. Then the ratio of the $\chi^2$'s for the vector to scalar degeneracy breaking is

$$\frac{\chi^2_{--}}{\chi^2_{++}} = \left(\frac{\sigma_{\beta,+}}{\sigma_\omega}\Omega t_e \frac{\delta\beta^{--}}{\delta\beta^{++}}\right)^2 \sim \left(3\frac{t_e}{22\,\text{days}}\frac{\delta\beta^{--}}{\delta\beta^{++}}\right)^2. \qquad (2.16)$$

Again, while it is not impossible that $\delta\beta^{--}$ be much smaller than the typical values of $\delta\beta^{++}$, if it is too small, then the $\pm\Delta\beta_-$ solutions are so similar that it is unimportant to break the degeneracy.

These qualitative conclusions are born out quantitatively in the remainder of the paper.

## 3. Halo Model and Monte Carlo

The precise structure of the Milky Way halo is still unknown. In the interests of simplicity, we perform our simulations with the simplest existing model: the isothermal sphere. We discuss the effects of using other models below.

We fix the macho mass $M$ for each simulation and take the macho mass density $\rho$ to be,

$$\rho(r) = \frac{\rho_0}{1 + (r/a_0)^2}, \tag{3.1}$$

where $r$ is the distance to the Galactic center (assumed to be 8 kpc at the Sun) and $a_0 = 5.6$ kpc is the core radius. The results do not depend on the value of $\rho_0$. The macho velocity distribution is taken to be an isotropic Gaussian in the Galactic frame. The distribution of the transverse speed $v$ (in the plane perpendicular to the line of sight) is

$$\frac{dN}{dv} \propto f(v) = v \, \exp\left(-\frac{v^2}{2\sigma^2}\right), \tag{3.2}$$

where $\sqrt{2}\sigma = 210 \, \mathrm{km \, s^{-1}}$ is normalized to the local Galactic rotation speed.

Although their effect is small, we take into account the motion of both the Sun and the LMC in the Galactic frame. Both are $\sim 230 \, \mathrm{km \, s^{-1}}$ (Jones et al. 1994). These motions induce a motion of the line of sight, and a small correlation between the probability distributions of the position and the speed of the machos. We neglect this correlation. The contribution to the event rate of a macho scales as $\rho(r) \, r_e \, v \propto \rho(r) \sqrt{z(1-z)} \, v$. Hence we choose randomly the macho position in the line of sight with the probability distribution $\sqrt{z(1-z)} \, \rho(r)$, and the macho speed $v$ with the probability distribution $v \, f(v)$. The angle between the direction of this velocity and a fixed direction in the plane perpendicular to the line of sight is chosen randomly from a uniform distribution between 0 and $2\pi$.

The impact parameter $\beta$ is either fixed, or chosen randomly with a flat distribution between 0 and $\beta_{max}$. We fix the Earth–satellite separation, the precisions of the ground and the satellite photometries, and the frequency of observations. To simulate the experiment, we assume that the first satellite observation starts when the source enters the Einstein ring as seen from the Earth, and we assume that both sites stop observing when the source–lens separation is greater than $3r_e$ for both the Earth and the satellite. The events where the minimum source–lens separation for the satellite $\beta' > 1.5r_e$ are assumed to be unresolvable.

We then calculate the four possible solutions for the projected velocity $\tilde{\mathbf{v}}$, and their intrinsic errors. For each solution, a $\chi^2$ value quantifies its compatibility with the degeneracy breaking constraint [eq. (2.7)]. Those solutions with $\chi^2 \leq 9$ are retained, and those with $\chi^2 > 9$ are excluded. If the fractional difference between all the allowed solutions and the true value is less than 20%, and if the intrinsic fractional errors (at $1\sigma$) of the allowed solutions are less than 20%, we consider that the degeneracy is broken and that we know the projected speed with an error $\lesssim 20\%$. (As we discuss analytically in § 2 and numerically in § 4, the total errors are dominated by the degeneracy breaking, so the total error is still typically $\lesssim 20\%$.) For each set of parameters, the simulation is repeated 5000 times, yielding Monte Carlo estimates with statistical errors less than 1%.

The effects of using a different model can be understood qualitatively as follows. For models (such as a flattened halo model) in which the machos are a factor $\zeta$ closer to the Earth, the effect is to make the Einstein ring smaller by $\zeta^{1/2}$. This is very much the same as reducing the mass by a factor $\zeta$ and the effect can be judged by examining Figure 3 below. For models with slower macho velocities, the effect is very similar to having larger Earth-satellite relative velocities. This tends to increase the accuracy as can also be seen in Figure 3 below. The principal model for which one would expect substantially different velocities is a truncated halo model, or more generally, a halo whose density falls $r^{-n}$, $(n > 3)$. For such models, the velocity dispersion is lower than for an isothermal sphere by $\sqrt{2/3}$ (see problem 4.9 of Binney & Tremaine 1987). One also expects the machos to be closer to the Earth in such models, so there is little net effect on the degeneracy breaking fraction.

## 4. Results I: The Speed Degeneracy

The simulated degeneracy breaking fraction (DBF) depends on six parameters: the photometric precisions $\sigma_E$ and $\sigma_S$ of the Earth and satellite measurements, the number of observations done per Einstein radius crossing time $n_{obs}$, the maximum value $\beta_{max}$ of $\beta$, the fixed mass $M$ of the machos, and the Earth–satellite separation $d_{sat}$. (Because we assume the satellite is in a heliocentric Earth-like orbit, we measure this separation in time, one month being equivalent to an angle of $2\pi/12$.) An observing rate $n_{obs} = 20$ corresponds to one observation per 0.5 to 4 days for the events reported to date which are typically $t_e$ ~10–75 days. We therefore consider this to be a realistic observing rate. Fiducial parameters for the whole simulation are $\sigma_E = 3\%$, $\sigma_S = 4\%$, $n_{obs} = 20$, $d_{sat} = 2$ months, $M = 0.1 M_\odot$, and $\beta_{max} = 0.7$; in this case we find DBF = 80%. Here, we present curves representing the DBF when 1 or 2 out of the 6 parameters are changed.

Figure 1 presents the DBF against $n_{obs}$ for photometric precisions $\sigma_E$ and $\sigma_S$ of respectively 4% and 5%, 3% and 4%, and 2% and 3%. To resolve 75% of the events, 25 observations per Einstein ring crossing time are necessary in case of 4% and 5% photometry, 15 for 3% and 4% photometry, and only 8 observations are sufficient for 2% and 3% photometry. Hence it is possible to break the two fold ambiguity for a large majority of the events with realistic observing conditions.

Figure 2 presents the DBF against $\beta$, which is not randomly chosen but constant for each point of the curve. The events where $\beta \geq 0.7$ are much harder to resolve. They should therefore be separated in our statistical analysis. For $0 \leq \beta \leq 0.7$, the mean DBF equals 80%; however for $0.7 \leq \beta \leq 1$, it equals 15%.

Figure 3 presents the DBF against the macho mass $M$, which is constant for each point of the figure, for five different Earth–satellite separations. The observed microlensing events reported up to now correspond very probably to masses in the range 0.01–1 $M_\odot$. We focus on these masses in the figure. We show curves for $d_{sat}$ = 0.5, 1, 2, 4, and 6 months because one possibility is to launch a satellite which would drift away from the Earth at a roughly constant rate of several months per

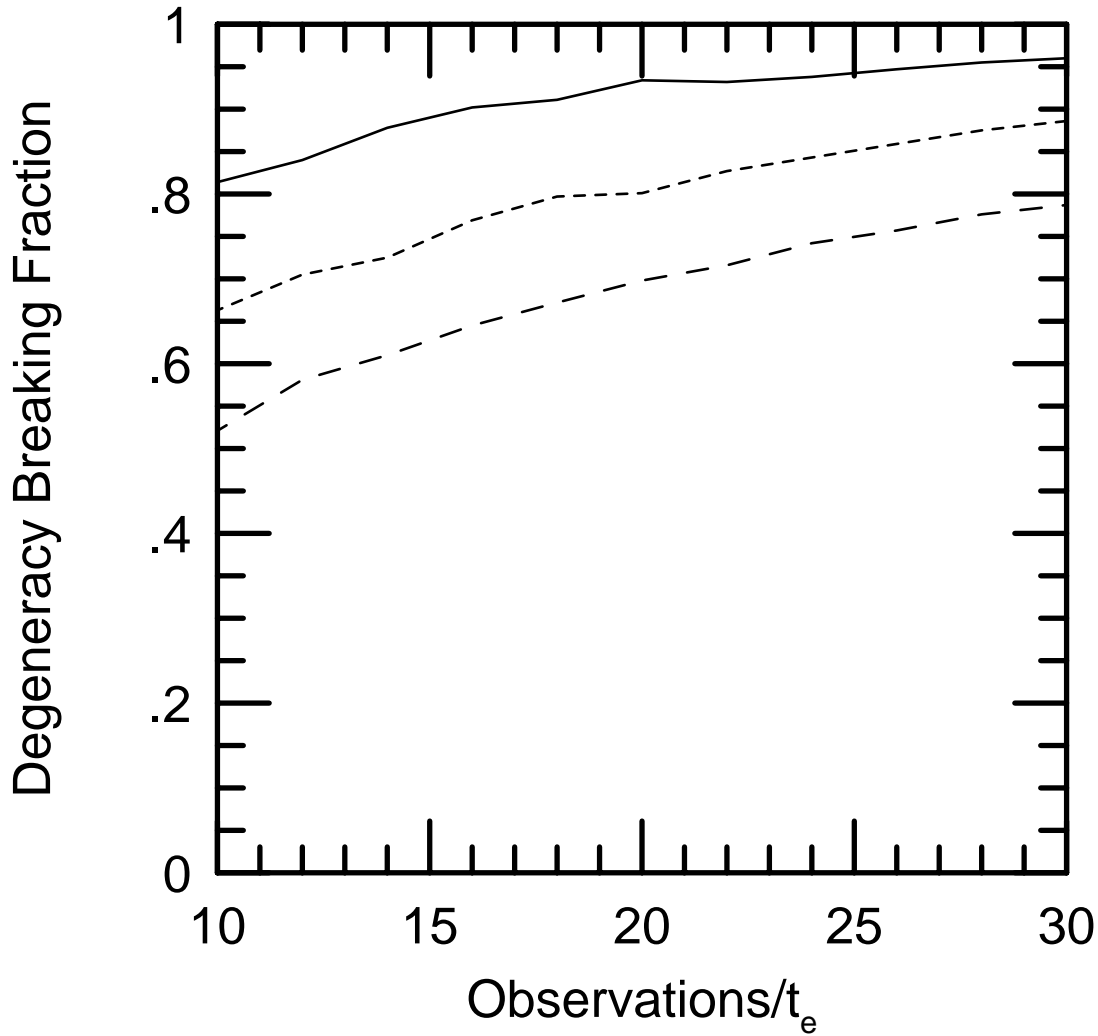

Figure 1. Degeneracy Breaking Fraction as a function of observation rate $n_{obs}$ (number of observations per Einstein radius crossing time) for three levels of fractional photometric accuracy from the Earth ($\sigma_E$) and the satellite ($\sigma_S$). Shown are ($\sigma_E, \sigma_S$) = (2%,3%) (*solid curve*), (3%,4%) (*short dashed curve*), and (4%,5%) (*long dashed curve*). Other parameters are held fixed at $M = 0.1 M_\odot$, $d_{sat} = 2$ months (1 AU), and $\beta_{max} = 0.7$.

year. The set of curves then represents the evolution of the satellite efficiency over time.

The two fold ambiguity in $\tilde{v}$ can be most easily broken when $d_{sat}$ is close to the projected Einstein radius $\tilde{r}_e$: when $d_{sat} \ll \tilde{r}_e$, the event looks very similar from the Earth and the satellite, so the satellite observations are of little help; when $d_{sat} \gg \tilde{r}_e$, the probability that the satellite will not observe the event is high. Since $\tilde{r}_e \propto M^{1/2}$, there is an optimal macho mass for each $d_{sat}$, the mass for which the DBF is a maximum on the curves, and this optimal mass increases with $d_{sat}$. Since it is the relative velocity of the Earth and the satellite that allows one to break the degeneracy, and since this velocity increases with $d_{sat}$, the maximum DBF also increases with $d_{sat}$.

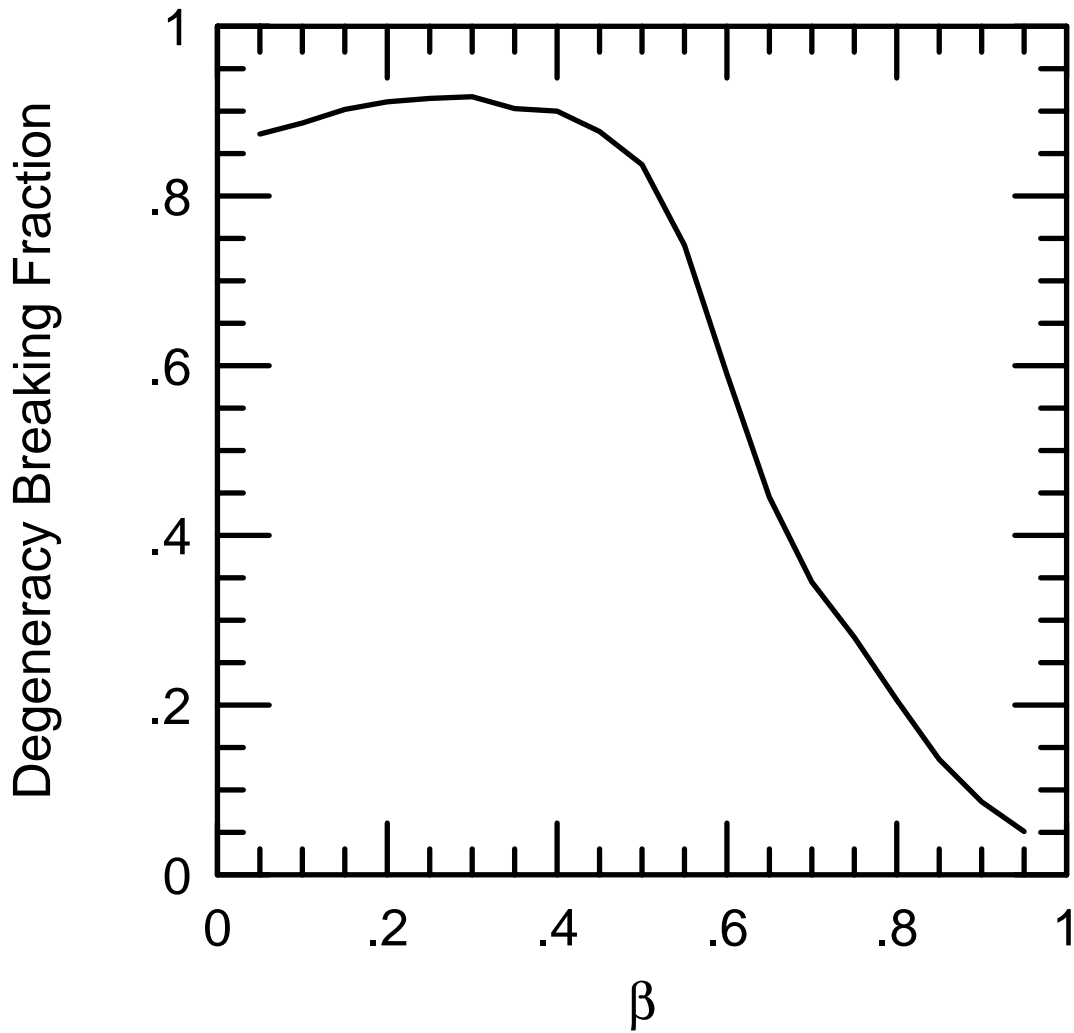

Figure 2. Degeneracy Breaking Fraction as a function of dimensionless impact parameter $\beta$. Other parameters are held fixed at $n_{obs} = 20$, $\sigma_E = 3\%$, $\sigma_S = 4\%$, $M = 0.1\,M_\odot$, and $d_{sat} = 2$ months.

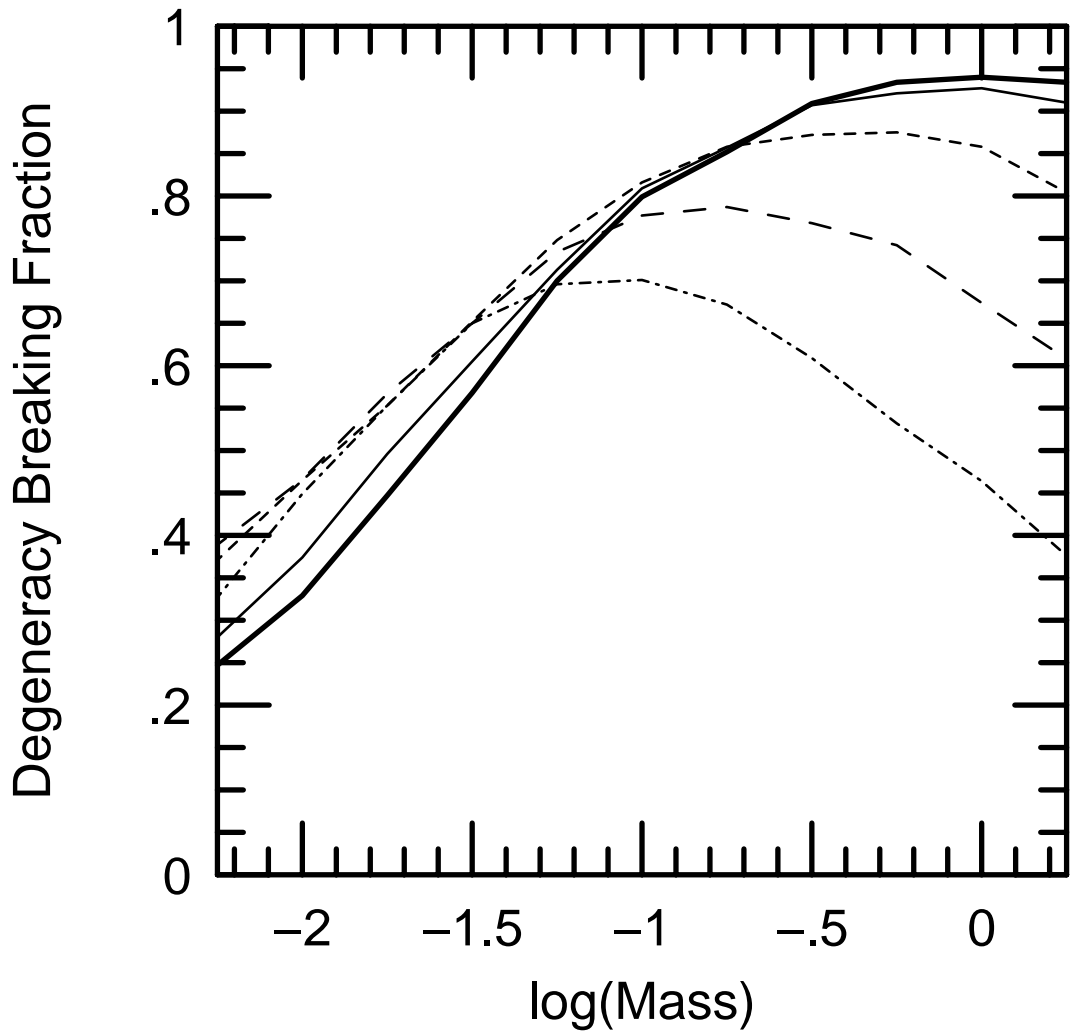

Figure 3. Degeneracy Breaking Fraction as a function of $\log_{10}$ of the macho mass in solar units for five values of $d_{sat}$, the distance between the Earth and the satellite. Shown are $d_{sat} = 0.5$ months (0.26 AU) (*dot-dashed curve*), 1 month (0.52 AU) (*long dashed curve*), 2 months (1 AU) (*short dashed curve*), 4 months (1.73 AU) (*solid curve*), and 6 months (2 AU) (*bold solid curve*). Other parameters are held fixed at $n_{obs} = 20$, $\sigma_E = 3\%$, $\sigma_S = 4\%$, $M = 0.1\,M_\odot$, and $\beta_{max} = 0.7$.

The satellite efficiency depends weakly on $d_{sat}$ at $\sim 0.01\,M_\odot$, and increases quickly with $d_{sat}$ in the range 0.1–1 $M_\odot$. It is important to notice that one always benefits by increasing $d_{sat}$. Globally, the DBF remains higher than 40% and most of the DBF values are in the range 60–90%.

For very large masses (beyond the window shown in Fig. 3), the events are also well resolved: for $M = 10 M_\odot$, the DBF is between 50% and 80% when $d_{sat}$ is larger than 1.5 months. The results are less satisfactory for very small masses: when $M = 0.001 M_\odot$, the best DBF values are in the range 20–30%, obtained when $d_{sat} \leq 2$ months.

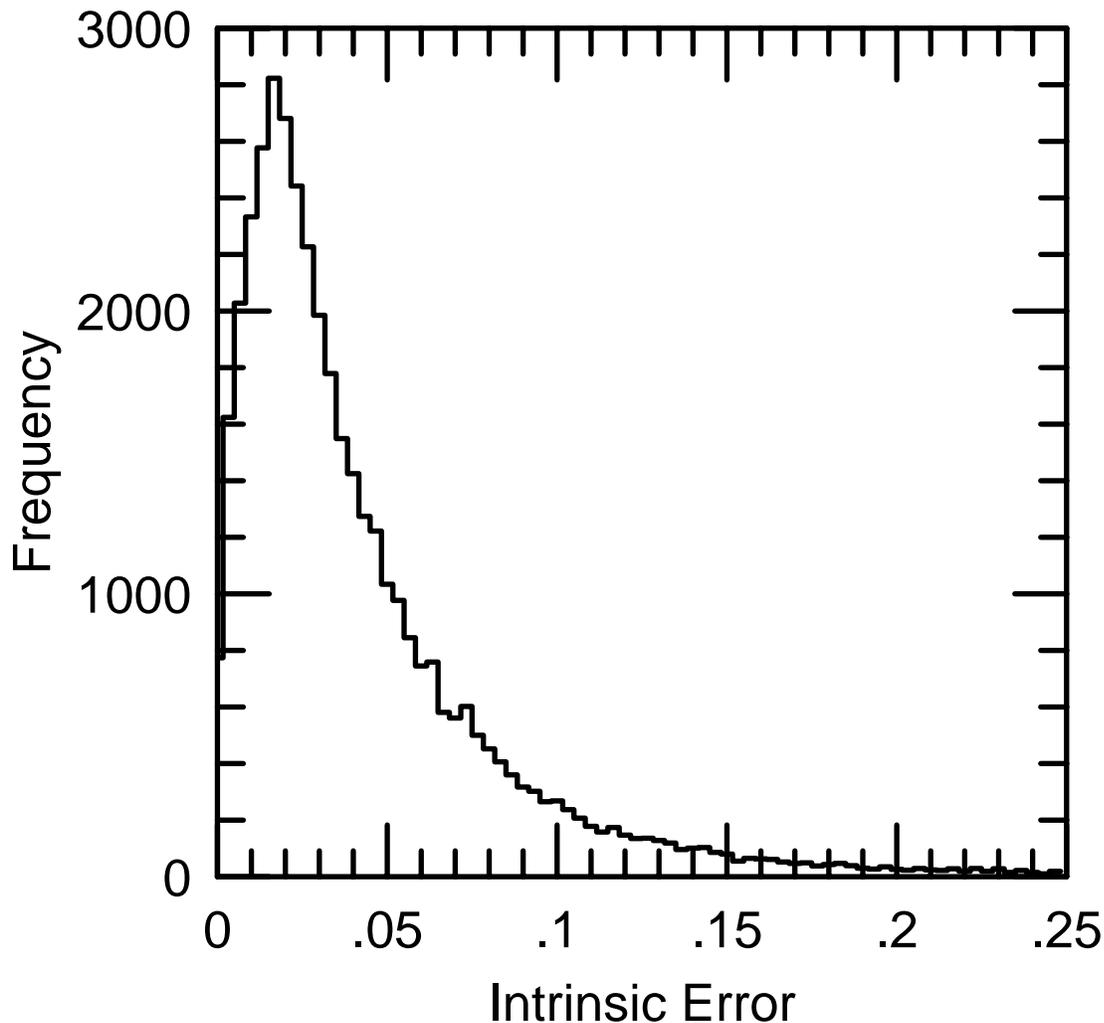

Figure 4. Number of trials as a function of the largest intrinsic error in $\tilde{v}$ [see eq. (2.12)] of any allowed solution. Trials where there were two allowed solutions which differed by more than 20% are excluded from this plot. Parameters are held fixed at $n_{obs} = 20$, $\sigma_E = 3\%$, $\sigma_S = 4\%$, $M = 0.1 M_\odot$, $\beta_{max} = 0.7$, and $d_{sat} = 2$ months.

In § 2, we argued analytically that the uncertainty in determining $\tilde{v}$ is dominated by the problem of resolving the degeneracy between $\Delta\beta_\pm$ solutions and that the intrinsic error around each solution plays only a minor role. Figure 4 provides a numerical justification for this claim. It shows the intrinsic error for all cases where either a) there was only one allowed solution or b) there were two or more allowed solutions, but these differed by $< 20\%$. Note that the intrinsic errors are typically $\sim 2\%$ and are rarely greater than $10\%$.

## 5. Results II: The Velocity Degeneracy

As we emphasized in the introduction, the most important requirement for a parallax satellite is that it break the two fold degeneracy in projected speed $\tilde{v}$, because this allows one to determine to which component of the Galaxy the detected machos belong. In the previous section we determined the observational requirements to meet this goal.

Nevertheless, the additional information contained in the projected velocity $\tilde{\mathbf{v}}$ can also be important. For example, if the machos lie in a non-rotating isotropic halo, then their velocities should be characterized by a projected asymmetric drift $\sim 100\,\mathrm{km\,s^{-1}}$. On the other hand a rotating halo might have little or no asymmetric drift.

In this section we test how well the velocities are measured given the observational parameters ($n_{obs} = 20$, $\sigma_E = 3\%$, $\sigma_S = 4\%$) required to break the speed degeneracy as determined in the previous section. For degenerate solutions, we define the velocity error as the magnitude of the difference between the true and the allowed solution, divided by the true speed. We evaluate the intrinsic errors of each solution using equation (2.11). As in the previous section, we demand that each type of error is less than $20\%$. The results are presented in Figure 5 and can be directly compared with Figure 3. We find that the DBF for velocities is only slightly reduced from the DBF for speeds. For example, for $M = 0.1\,M_\odot$ and $d_{sat} = 2$ months, the respective fractions are $67\%$ and $80\%$. In other words, it should be possible to extract substantial information about the directions of the machos as well as their speeds.

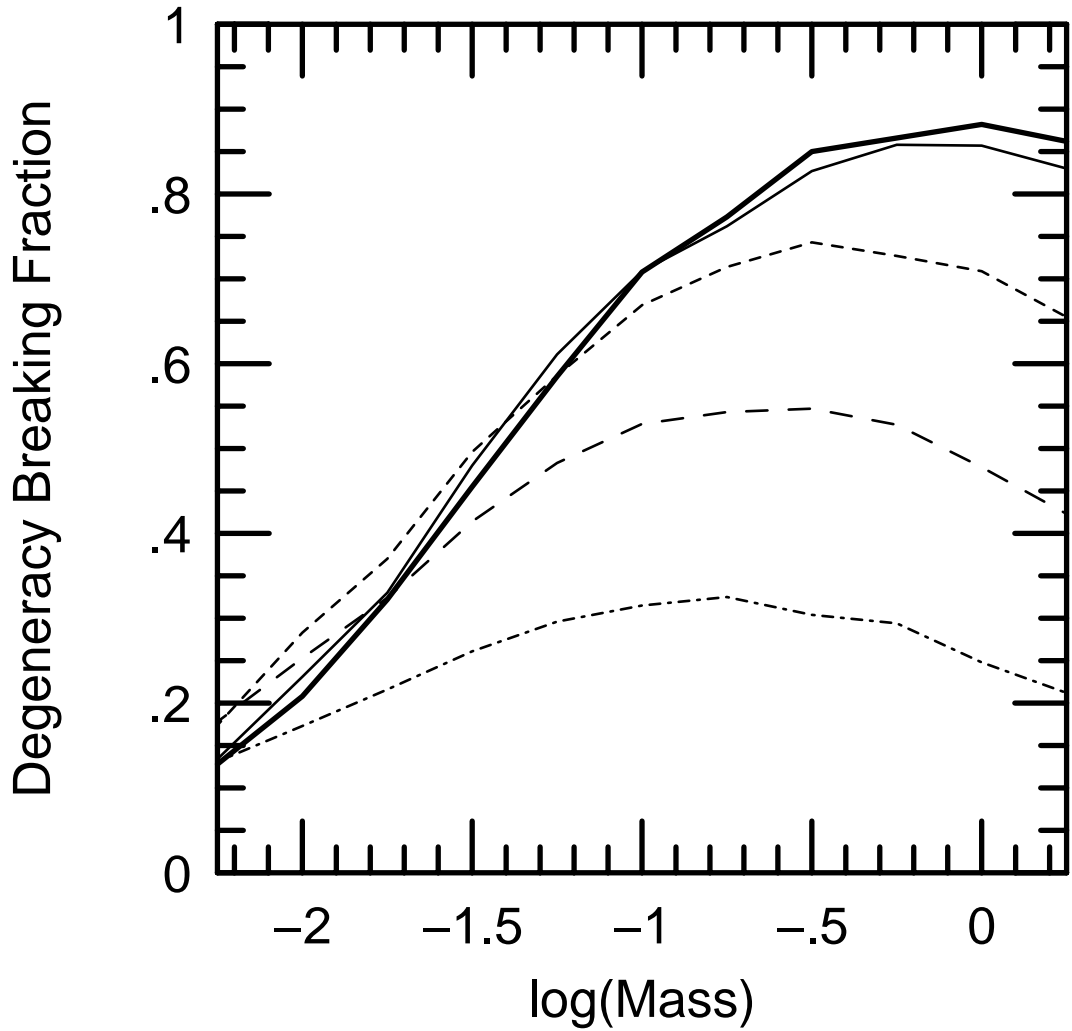

Figure 5.  Vector Degeneracy Breaking Fraction as a function of $\log_{10}$ of the macho mass in solar units for five values of $d_{sat}$. Symbols are the same as in Fig. 3. Vector degeneracy breaking shown here requires that the projected velocity $\tilde{\mathbf{v}}$ be determined to 20% whereas the DBF shown in Fig. 3 requires only that $\tilde{v}$ be determined to this accuracy.

## 6. Conclusions

We find that in feasible observing conditions, with a ground photometric precision of 3% and a satellite photometric precision of 4%, and when the Earth–satellite separation is larger than one month (i.e., 0.5 AU) it is possible to measure the projected speed $\tilde{v}$ of events with $\lesssim 20\%$ accuracy for at least 70% of LMC microlensing events similar to the ones reported by the EROS and MACHO collaborations.

Under the same conditions we find that the projected velocity $\tilde{\mathbf{v}}$ can be resolved at the same level at least 50% of the time.

In these analyses we have deliberately excluded events with impact parameters $\beta > 0.7$. Such events are not at present being discovered by EROS and MACHO because they are essentially excluded by the event-selection criteria. If such events are discovered in the future, they should be excluded from satellite follow-up observations because, as we have shown, the probability that the reduced speed can be measured is very small (15%).

We find that it is generally advantageous to have larger rather than smaller Earth–satellite separations: there is greater precision for masses $M \sim 0.1$–$1\,M_\odot$ and there is nearly no penalty for masses $M \sim 0.01\,M_\odot$. Since it is expected that large separations are also favored for observations toward the bulge (Gould 1995a; Gaudi & Gould 1995), large separations appear to be favored by all considerations.

**Acknowledgments** : We would like to thank the EROS collaboration for its hospitality. In particular, advice and comments from A. Milsztajn and J. Rich are gratefully acknowledged. This work was supported in part by grant AST 94-20746 from the NSF and in part by grant NAG5-2864 from NASA.